\documentclass[prd,twocolumn, tightten, aps, nofootinbib, showpacs, preprintnumbers]{revtex4}
\begin{document}
\preprint{USM-TH-212}
\title{Properties of noncommutative axionic electrodynamics}
\author{Patricio Gaete}
\email{patricio.gaete@usm.cl}
\author{Iv\'an  Schmidt}
\email{ivan.schmidt@usm.cl}
\affiliation{Center of Subatomic Studies and
Departmento de F\'{\i}sica\\ Universidad T\'ecnica
Federico Santa Mar\'{\i}a, Valpara\'{\i}so, Chile}
\date{\today}

\begin{abstract}
Using the gauge-invariant but path-dependent variables formalism, we
compute the static quantum potential for noncommutative axionic
electrodynamics, and find a radically different result than the
corresponding commutative case. We explicitly show that the static
potential profile is analogous to that encountered in both
non-Abelian axionic electrodynamics and in Yang-Mills theory with
spontaneous symmetry breaking of scale symmetry.
\end{abstract}
\maketitle

The coupling of axion-like particles (or ''axions'') with photons in
the presence of an external background electromagnetic field and its
physical consequences such as vacuum birefringence and vacuum
dichroism have been the object of intensive investigations by many
authors
\cite{Maiani,Sikivie,Stodolsky,Grifols,Jain,Masso,Heyl,Raffelt,Ringwald}.
Moreover, as is well known, this subject has received increasing
attention after recent results of the PVLAS collaboration
\cite{Zavattini}, in which dichroism \cite{Zavattini} and
birefringence \cite{Cantatore} were observed for a linearly
polarized laser beam after it traverses an external magnetic field.
We recall that these effects can be qualitatively understood by the
existence of light pseudoscalars bosons $\phi$ (''axions''), with a
coupling to two photons. In this case, the corresponding term in the
effective Lagrangian  has the form $ {\cal L}_I  = -
{\raise0.7ex\hbox{$1$} \!\mathord{\left/
 {\vphantom {1 {4M}}}\right.\kern-\nulldelimiterspace}
\!\lower0.7ex\hbox{${4M}$}}F_{\mu \nu } {\widetilde F}^{\mu \nu
}\phi$, where $ {\widetilde F}^{\mu \nu }  = {\raise0.7ex\hbox{$1$}
\!\mathord{\left/{\vphantom {1 2}}\right.\kern-\nulldelimiterspace}
\!\lower0.7ex\hbox{$2$}}\varepsilon _{\mu \nu \lambda \rho }
F^{\lambda \rho }$. Certainly, if the PVLAS results are confirmed by
further experimental data, it would signal new physics containing
very light bosons \cite{Bibber}. Meanwhile, alternatives models have
been suggested in order to circumvent the discrepancy among the
results of the PVLAS experiment with both astrophysical bounds and
the results of the CAST collaboration \cite{Masso,Antoniadis}. For
example, in Ref. \cite{Antoniadis} the existence of a new light
vector field ( rather than an axion field) was considered, which
interacts with the photon via Chern-Simons-like terms. Another
possible scenario is the existence of millicharged particles
\cite{Gies}. In addition, the study of space-time non-commutativity
on light propagation in a background electromagnetic field has also
attracted interest in order to account the results reported by PVLAS
collaboration \cite{Sheik,Chatillon}. In particular, the
non-commutative axionic electrodynamics version has been considered
\cite{Chatillon} within the framework of the Lorentz violating
extension of QED \cite{Kostelecky}.

On the other hand, axionic electrodynamics experiences mass
generation due to the breaking of rotational invariance induced by a
classical background configuration of the gauge field strength
\cite{Spallucci}, and leads to confining potentials in the presence
of nontrivial constant expectation values for the gauge field
strength $F_{\mu \nu}$ \cite{GaeteGuen}. In fact, in the case of a
constant electric field strength expectation value the static
potential remains Coulombic, while in the case of a constant
magnetic field strength expectation value the potential energy is
the sum of a Yukawa and a linear potential, leading to the
confinement of static charges. In passing we note that the
distinction between the apparently related phenomena of screening
and confinement is of considerable importance in our present
understanding of gauge theories. We also mention that the magnetic
character of the field strength expectation value needed to obtain
confinement is in agreement  with the current chromo-magnetic
picture of the $QCD$ vacuum \cite{Savvidy}. Interestingly, similar
results have been obtained in the context of the dual
Ginzburg-Landau theory \cite{Suganuma}, as well as for a theory of
antisymmetric tensor fields that results from the condensation of
topological defects as a consequence of the Julia-Toulouse mechanism
\cite{GaeteW}. It must be clear from this discussion that the
interaction potential between static charges is an object of utmost
importance, which has a variety of applications. For instance, it
provides a useful framework for studying aspects of duality in
Maxwell-like three-dimensional models induced by the condensation of
topological defects driven by quantum fluctuations \cite{GaeteWot}.
Also, it has proven to be a key tool to analyze the
equivalence/nonequivalence between the $\theta$-expanded version of
the noncommutative U(1) gauge theory and the Born-Infeld action up
to order $F^{3}$ \cite{GaeteSchmidt}.

Inspired by these observations, the purpose of this note is to
further elaborate on the physical content of noncommutative axionic
electrodynamics \cite{Chatillon}. We examine another aspect of this
theory, namely, the effect of the space-time noncommutativity on a
physical observable. To do this, we will work out the static
potential for the theory under consideration by using the
gauge-invariant but path-dependent variables formalism, which is
alternative to the Wilson loop approach. Our treatment is fully
non-perturbative for the pseudoscalar field. As a result, we obtain
a Cornell-like potential which displays a marked departure of a
qualitative nature from the results of the commutative case
\cite{GaeteGuen} at large distances, and which clearly shows the key
role played by the noncommutative space in transforming the Yukawa
potential into the Coulombic one. Incidentally, the above static
potential profile is analogous to that encountered in both
non-Abelian axionic electrodynamics \cite{GaeteSpall} and in
Yang-Mills theory with spontaneous symmetry breaking of scale
symmetry \cite{GaeteSpall2}. In this way we establish a new
connection between noncommutative and non-Abelian effective
theories. The above connections are of interest from the point of
view of providing unifications among diverse models as well as
exploiting the equivalence in explicit calculations, as we are going
to show. Moreover, our work could be of interest for searching for
small violations of the Lorentz symmetry along the lines of Ref.
\cite{GaeteWprd}.

We now examine the interaction energy between static point-like
sources for noncommutative axionic electrodynamics. To do this, we
will compute the expectation value of the energy operator $H$ in the
physical state $|\Phi\rangle$ describing the sources, which we will
denote by $ {\langle H\rangle}_\Phi$. However, before going to the
derivation of the interaction energy, we will describe very briefly
the model under consideration. The initial point of our analysis is
the Lagrangian density:
\begin{equation}
{\cal L} = {\cal L}_{LV - QED}  + {\cal L}_{Axion}, \label{NCAE1}
\end{equation}
where ${\cal L}_{LV - QED}  =  - \frac{1}{4}F_{\mu \nu } F^{\mu \nu
} - \frac{1}{4}\left( {k_F } \right)_{\mu \nu \rho \sigma } F^{\mu
\nu } F^{\rho \sigma }  + \frac{1}{2}\left( {k_{AF} } \right)^\mu
\varepsilon _{\mu \nu \rho \sigma } A^\nu  F^{\rho \sigma }$
represents the Lorentz-violating extension of QED \cite{Kostelecky},
and ${\cal L}_{Axion}  = \frac{1}{2}\left( {\partial \phi }
\right)^2 - \frac{{m^2 }}{2}\phi ^2  + \frac{\phi }{{8M}}\varepsilon
_{\mu \nu \rho \sigma } F^{\mu \nu } F^{\rho \sigma }$ describes the
coupling of a pseudoscalar $\phi$ to electromagnetism. Next, the
corresponding noncommutative version of (\ref{NCAE1}) is induced by
the Moyal product $f * g = f\exp \left( {\frac{i}{2}\overleftarrow
\partial  _\mu  \theta ^{\mu \nu } \overrightarrow \partial  _\nu  }
\right)g$, where $\theta ^{\mu \nu}$ is a real constant
antisymmetric tensor. As a consequence, the field strength
${\widehat{F}}_{\mu\nu}$ in a noncommutative space-time is given by
$\widehat F_{\mu \nu }  = \partial _\mu \widehat A_\nu - \partial
_\nu  \widehat A_\mu   - ig\left[ {\widehat A_\mu ,\widehat A_\nu }
\right]$  and the bracket denotes a Moyal commutator. This then
implies that the theory is invariant under a non conventional gauge
transformation $\delta \widehat A_\mu   =
\partial _\mu  \widehat \lambda  - ig\left[ {\widehat A_\mu  ,\widehat
\lambda }\right]$.  Now we can exploit the Seiberg-Witten map
$\widehat A_\mu   = A_\mu   - \frac{1}{2}\theta ^{\alpha \beta }
A_\alpha  \left( {\partial _\beta  A_\mu   + F_{\beta \mu } }
\right)$, to get a gauge field $A_{\mu}$ with the ordinary gauge
transformation and a Lagrangian density written in terms of the
conventional field strength. Accordingly, after splitting
$F_{\mu\nu}$ in the sum of a classical background $\left\langle
{F_{\mu \nu } } \right\rangle$ and a small fluctuation $f_{\mu \nu }
=\partial _\mu A_\nu -\partial _\nu A_\mu$, the corresponding
Lagrangian density to leading order in $\theta$  ( with
$\theta^{0i}=0$ and $ \left( {k_{AF} } \right)^{\mu} = 0$) is given
by \cite{Chatillon}:
\begin{eqnarray}
{\cal L} &=&  - \frac{1}{4}\left( {1 - \frac{g}{2}\theta ^{ij}
\left\langle {F_{ij} } \right\rangle } \right)f^{\mu \nu } f_{\mu
\nu }  + \frac{g}{4}\theta ^{ij} f_{ij} \left\langle {F_{kl} }
\right\rangle f^{kl}   \nonumber \\
&-& \frac{g}{2}\theta _{ij} \left\langle
{F_{kl} } \right\rangle f^{ki} f^{lj}  - g\theta _{ij} \left\langle
{F^{ki} } \right\rangle f_{\mu k} f^{j\mu }  +
 \frac{1}{2}\left( {\partial \phi } \right)^2  \nonumber \\
&-& \frac{{m^2 }}{2}
\phi ^2  + \frac{1}{{4M}}\varepsilon ^{\rho \sigma \mu \nu }
\left\langle {F_{\rho \sigma } } \right\rangle f_{\mu \nu } \phi .
\label{NCAE5}
\end{eqnarray}
Throughout, $M$ and $g$ are the photon-axion and U(1) gauge coupling
respectively. To get the last equation we used $\left( {k_F } \right)_{\mu
 \nu \rho \sigma }  = \frac{1}{8}\left( {T_{\left[ {\mu \nu } \right]\left[
  {\rho \sigma } \right]}  + \mu \nu  \leftrightarrow \rho \sigma } \right)$,
where $ T_{\mu \nu \rho \sigma }  \equiv  - \frac{1}{2}g\theta
^{\alpha \beta } \left\langle {F_{\alpha \beta } } \right\rangle
\eta _{\mu \rho } \eta _{\nu \sigma }  - g\theta _{\mu \nu }
\left\langle {F_{\rho \sigma } } \right\rangle  + 4g\theta _{\alpha
\nu } \left\langle {F_\rho ^\alpha  } \right\rangle  + 2g\theta
_{\nu \sigma } \left\langle F \right\rangle _{\mu \rho }$, and $
T_{\left[ {\mu \nu } \right]}  \equiv T_{\mu \nu }  - T_{\nu \mu }$.
Here $(\mu ,\nu, \rho, \sigma  = 0,1,2,3)$ and $(i, j, l, k =
1,2,3)$.

Following our earlier procedure \cite{GaeteGuen}, integrating out
the $\phi$ field induces an effective theory for the $A_{\mu}$
field. Once this is done, we arrive at the following effective
Lagrangian density:
\begin{eqnarray}
{\cal L} &=&  - \frac{1}{4}\left( {1 - \frac{g}{2}\theta ^{ij}
\left\langle {F_{ij} } \right\rangle } \right)f^{\mu \nu } f_{\mu \nu }
+ \frac{g}{4}\theta ^{ij} f_{ij} \left\langle {F_{kl} } \right\rangle f^{kl}
\nonumber  \\
&-& \frac{g}{2}\theta _{ij} \left\langle {F_{kl} } \right\rangle f^{ki} f^{lj}
- g\theta _{ij} \left\langle {F^{ki} } \right\rangle f_{\mu k} f^{j\mu } \nonumber  \\
 &+& \frac{1}{{32M^2 }}v^{\mu \nu } f_{\mu \nu } \frac{1}{{\Delta  + m^2 }}
v^{\gamma \delta } f_{\gamma \delta }. \label{NCAE10}
\end{eqnarray}
Here we have simplified our notation by setting $\varepsilon
 ^{\mu \nu \alpha \beta } \left\langle{F_{\mu \nu } } \right\rangle
\equiv v^{\alpha \beta }$ and $\varepsilon ^{\rho \sigma \gamma
 \delta } \left\langle {F_{\rho \sigma } } \right\rangle
 \equiv v^{\gamma \delta }$. One immediately sees that this expression has
a remarkable similarity to the corresponding commutative effective
Lagrangian density. This common feature is our main motivation for
studying the role of the noncommutative space on the interaction
energy. With this in view, we now proceed to calculate the
interaction energy in the $v^{0i} \ne 0$ and $v^{ij}=0$ case
(referred to as the magnetic one in what follows). Using this in
(\ref{NCAE10}) we then obtain
\begin{eqnarray}
{\cal L} &=&  - \frac{1}{4}\left( {1 - \frac{g}{2}\theta ^{ij}
\left\langle {F_{ij} } \right\rangle } \right)f^{\mu \nu } f_{\mu \nu }
+ \frac{g}{4}\theta ^{ij} f_{ij} \left\langle {F_{kl} } \right\rangle f^{kl}
\nonumber  \\
&-&  \frac{g}{2}\theta _{ij} \left\langle {F_{kl} } \right\rangle f^{ki}
f^{lj}  - g\theta _{ij} \left\langle {F^{ki} } \right\rangle f_{\mu k} f^{j\mu }
\nonumber  \\
&+&  \frac{1}{{16M^2 }}v^{0i} f_{0i} \frac{1}{{\Delta  + m^2
}}v^{0k} f_{0k}. \label{NCAE15}
\end{eqnarray}
We observe that the limit $\theta\rightarrow0$ is well defined, and
leads to the corresponding commutative Lagrangian density. To obtain
the corresponding Hamiltonian, we must carry out the quantization of
this theory. The Hamiltonian analysis starts with the computation of
the canonical momenta  $ \Pi ^\mu   =  - \left( {1 -
\frac{g}{2}\theta ^{ij} \left\langle {F_{ij} }
 \right\rangle } \right)f^{0\mu }  - g\theta _{ij} \left\langle {F^{\mu i} }
\right\rangle f^{j0}  + g\theta ^{i\mu } \left\langle {F^{ki} }
\right\rangle f^{0k}  + \frac{1}{{16M^2 }}v^{0\mu } \frac{1}{{\Delta
+ m^2 }}v^{oi} f_{0i}$, which produces the usual
primary constraint  $\Pi^{0}=0$ and $\Pi _i  = \left( {1 -
\frac{g}{2}\theta ^{ij} \left\langle {F_{ij} } \right\rangle }
\right)f_{i0}  - g\left[ {\theta _{kj} \left\langle {F_{kj} }
\right\rangle  + \theta _{ki} \left\langle {F_{jk} } \right\rangle }
\right]f_{j0}  + \frac{1}{{16M^2 }}v_{i0} \frac{1}{{\Delta  + m^2
}}v_{j0} f_{j0}$. The canonical Hamiltonian to leading order in
$\theta$ and $\frac{1}{M^{2}}$ is then
\begin{eqnarray}
H_C & =& \int {d^3 x} \left\{ { - A^0 \left( {\partial _i \Pi ^i}
\right) + \frac{{\left[ {1 + g\left( {{\bf \theta} \cdot {\bf {\cal
B}} }
\right)} \right]{\bf \Pi} ^2 }}{2}} \right\} \nonumber  \\
 &+&  \int {d^3 x} \left\{ {\frac{{\left( {{\bf \theta} \cdot {\bf {\cal B}} }
\right)}}{2}{\bf B}^2  - \frac{{g\left( {{\bf \theta}\cdot {\bf B}} \right)}}
{2}{\bf {\cal B}} \cdot {\bf B}} \right\} \nonumber  \\
 &+& \int {d^3 x} \left\{ {g\theta _{ij} \left[ {\frac{{\left\langle {F_{kl} }
\right\rangle }}{2}f^{ki}  - \left\langle {F^{ki} } \right\rangle f_{lk} }
\right]f^{lj} } \right\} \nonumber \\
&+& \int {d^3 x} \frac{{g\left( {{\bf {\cal B}} } \right)^2 \left( {{\bf \theta}
\cdot {\bf {\cal B}} } \right)}}{{32M^2 }}\Pi ^i \left( {\Delta  + m^2 }
\right)^{ - 1} \Pi ^i  \nonumber  \\
 &-&\int {d^3 x} \frac{{\left( {{\bf {\cal B}} } \right)^2 \left[ {1 +
5g\left( {{\bf \theta}\cdot {\bf {\cal B}} } \right)} \right]}}{{32M^2 }}
\Pi ^i \left( {\Delta  + {\cal M}^2 } \right)^{ - 1} {\Pi}^{i} \nonumber, \\
\label{NCAE20}
\end{eqnarray}
where ${\cal M}^2 \equiv m^2 + \frac{{{\bf v}^2 }}{{16M^2 }} = m^2 +
\frac{{{\bf {\cal B}^2 }}}{{4M^2 }}$. Here ${\bf B}$ and ${\bf {\bf
{\cal B}}}$ represent the magnetic field and external magnetic
field, respectively. Requiring the primary constraint $ \Pi_0=0$ to
be preserved in time yields the secondary constraint $\Gamma_1
\left( x \right) \equiv \partial _i \Pi ^i=0$. But the time
stability of the secondary constraint does not induce further
constraints. Therefore, the extended Hamiltonian that generates
translations in time then reads $H = H_C + \int {d^3 }x\left( {c_0
\left( x \right)\Pi _0 \left( x \right) + c_1 \left( x\right)\Gamma
_1 \left( x \right)} \right)$, where $c_0 \left( x\right)$ and $c_1
\left( x \right)$ are the Lagrange multiplier fields. Moreover, it
is straightforward to see that $\dot{A}_0 \left( x \right)= \left[
{A_0 \left( x \right),H} \right] = c_0 \left( x \right)$, which is
an arbitrary function. Since $ \Pi^0 = 0$ always, neither $ A^0 $
nor $ \Pi^0 $ are of interest in describing the system and may be
discarded from the theory. Thus the Hamiltonian is now given as
\begin{eqnarray}
H_C & =& \int {d^3 x} \left\{ { c(x) \left( {\partial _i \Pi ^i}
\right) + \frac{{\left[ {1 + g\left( {{\bf \theta} \cdot {\bf {\cal
B}} }
\right)} \right]{\bf \Pi} ^2 }}{2}} \right\} \nonumber  \\
 &+&  \int {d^3 x} \left\{ {\frac{{\left( {{\bf \theta} \cdot {\bf {\cal B}} }
\right)}}{2}{\bf B}^2  - \frac{{g\left( {{\bf \theta}\cdot {\bf B}}
\right)}}
{2}{\bf {\cal B}} \cdot {\bf B}} \right\} \nonumber  \\
 &+& \int {d^3 x} \left\{ {g\theta _{ij} \left[ {\frac{{\left\langle {F_{kl} }
\right\rangle }}{2}f^{ki}  - \left\langle {F^{ki} } \right\rangle
f_{lk} }
\right]f^{lj} } \right\} \nonumber \\
&+& \int {d^3 x} \frac{{g\left( {{\bf {\cal B}} } \right)^2 \left(
{{\bf \theta} \cdot {\bf {\cal B}} } \right)}}{{32M^2 }}\Pi ^i
\left( {\Delta  + m^2 }
\right)^{ - 1} \Pi ^i  \nonumber  \\
 &-&\int {d^3 x} \frac{{\left( {{\bf {\cal B}} } \right)^2 \left[ {1 +
5g\left( {{\bf \theta}\cdot {\bf {\cal B}} } \right)}
\right]}}{{32M^2 }}
\Pi ^i \left( {\Delta  + {\cal M}^2 } \right)^{ - 1} {\Pi}^{i}, \nonumber \\
\label{NCAE25}
\end{eqnarray}
where $c(x) = c_1 (x) - A_0 (x)$.

In accordance with the Dirac method, we must fix the gauge. A
particularly convenient gauge fixing condition is
\begin{equation}
\Gamma _2 \left( x \right) \equiv \int\limits_{C_{\xi x} } {dz^\nu }
A_\nu \left( z \right) \equiv \int\limits_0^1 {d\lambda x^i } A_i
\left( {\lambda x} \right) = 0, \label{NCAE30}
\end{equation}
where  $\lambda$ $(0\leq \lambda\leq1)$ is the parameter describing
the spacelike straight path $ x^i = \xi ^i  + \lambda \left( {x -
\xi } \right)^i $, and $ \xi $ is a fixed point (reference point).
There is no essential loss of generality if we restrict our
considerations to $ \xi ^i=0 $. The choice (\ref{NCAE30}) leads to
the Poincar\'e gauge \cite{Pato,GaeteSprd}. Through this procedure,
we arrive at the only nonvanishing equal-time Dirac bracket for the
canonical variables
\begin{eqnarray}
\left\{ {A_i \left( x \right),\Pi ^j \left( y \right)} \right\}^ *
&=&\delta{ _i^j} \delta ^{\left( 3 \right)} \left( {x - y} \right)  \nonumber \\
&-& \partial _i^x \int\limits_0^1 {d\lambda x^j } \delta ^{\left( 3
\right)} \left( {\lambda x - y} \right). \label{NCAE35}
\end{eqnarray}

After achieving the quantization we may now proceed to calculate the
interaction energy in the model under consideration. As mentioned
above, we will work out the expectation value of the energy operator
$H$ in the physical state $|\Phi\rangle$. Now we recall that the
physical states $|\Phi\rangle$ are gauge-invariant \cite{Dirac2}. In
that case we consider the stringy gauge-invariant state
\begin{eqnarray}
\left| \Phi  \right\rangle & \equiv& \left| {\overline \Psi  \left(
\bf y \right)\Psi \left( {\bf y}\prime \right)} \right\rangle \nonumber \\
&=& \overline \psi \left( \bf y \right)\exp \left(
{iq\int\limits_{{\bf y}\prime}^{\bf y} {dz^i } A_i \left( z \right)}
\right)\psi \left({\bf y}\prime \right)\left| 0 \right\rangle,
\label{NCAE40}
\end{eqnarray}
where $\left| 0 \right\rangle$ is the physical vacuum state and the
line integral appearing in the above expression is along a spacelike
path starting at ${\bf y}\prime$ and ending at $\bf y$, on a fixed
time slice. The charged matter field together with the
electromagnetic cloud (dressing) which surrounds it, is given by
$\Psi \left( {\bf y} \right) = \exp \left( { - iq\int_{C_{{\bf \xi}
{\bf y}} } {dz^\mu A_\mu  (z)} } \right)\psi ({\bf y})$. Thanks to
our path choice, this physical fermion then becomes $\Psi \left(
{\bf y} \right) = \exp \left( { - iq\int_{\bf 0}^{\bf y} {dz^i  }
A_{i} (z)} \right)\psi ({\bf y})$. In other terms, each of the
states ($\left| \Phi  \right\rangle$) represents a
fermion-antifermion pair surrounded by a cloud of gauge fields to
maintain gauge invariance.

From the foregoing Hamiltonian structure we then easily verify that
\begin{eqnarray}
\Pi _i \left( x \right)\left| {\overline \Psi  \left( \bf y
\right)\Psi \left( {{\bf y}^ \prime  } \right)} \right\rangle  &=&
\overline \Psi  \left( \bf y \right)\Psi \left( {{\bf y}^ \prime }
\right)\Pi _i \left( x \right)\left| 0 \right\rangle  \nonumber \\
&+&  q\int_ {\bf y}^{{\bf y}^ \prime  } {dz_i \delta ^{\left( 3
\right)} \left( {\bf z - \bf x} \right)} \left| \Phi \right\rangle.
\label{NCAE45}
\end{eqnarray}
Having made this observation and since the fermions are taken to be
infinitely massive (static) we can substitute $\Delta$ by
$-\nabla^{2}$ in Eq. (\ref{NCAE25}). In such a case
$\left\langle H \right\rangle _\Phi$ reduces to
\begin{equation}
\left\langle H \right\rangle _\Phi   = \left\langle H \right\rangle
_0 + \left\langle H \right\rangle _\Phi ^{\left( 1 \right)}  +
\left\langle H \right\rangle _\Phi ^{\left( 2 \right)}  +
\left\langle H \right\rangle _\Phi ^{\left( 3 \right)},
\label{NCAE50}
\end{equation}
where $\left\langle H \right\rangle _0  = \left\langle 0
\right|H\left| 0 \right\rangle$, and the $\left\langle H \right\rangle
 _\Phi ^{\left( 1 \right)}$, $\left\langle H \right\rangle
  _\Phi ^{\left( 2 \right)}$ and $\left\langle H \right\rangle
   _\Phi ^{\left( 3 \right)}$ terms are given by
\begin{equation}
\left\langle H \right\rangle _\Phi ^{\left( 1 \right)} = \alpha
\left\langle \Phi  \right|\int {d^3 x} \Pi ^i \Pi ^i \left| \Phi
\right\rangle, \label{NCAE55a}
\end{equation}
\begin{equation}
\left\langle H \right\rangle _\Phi ^{\left( 2 \right)} = \beta
\left\langle \Phi  \right|\int {d^3 x} \Pi ^i (\nabla ^2  - m^2
)^{-1}\Pi ^i \left| \Phi  \right\rangle, \label{NCAE55b}
\end{equation}

\begin{equation}
\left\langle H \right\rangle _\Phi ^{\left( 3 \right)}= \gamma
\left\langle \Phi  \right|\int {d^3 x} \Pi ^i  (\nabla ^2  - {\cal
M}^2 )^{-1}\Pi ^i \left| \Phi  \right\rangle, \label{NCAE55c}
\end{equation}
with $\alpha \equiv \frac{{\left[ {1 + g\left( {{\bf \theta}\cdot
{\bf {\cal B}}} \right)} \right]}}{2}$, $\beta \equiv \frac{{g\left(
{\bf {\cal B}} \right)^2 \left( {{\bf \theta}\cdot {\bf {\cal B}}}
\right)}}{{32M^2 }}$ and $\gamma \equiv \frac{{\left( {\bf {\cal B}}
\right)^2 \left[ {1 + 5g\left( {{\bf \theta}\cdot {\bf {\cal B}}}
\right)} \right]}}{{32M^2 }}$. Using Eq.(\ref{NCAE45}), the $\left\langle H \right\rangle
 _\Phi ^{\left( 1 \right)}$, $\left\langle H \right\rangle
  _\Phi ^{\left( 2 \right)}$ and $\left\langle H \right\rangle
   _\Phi ^{\left( 3 \right)}$ terms can be rewritten as
\begin{equation}
\left\langle H \right\rangle _\Phi ^{\left( 1 \right)}  = q^2 \alpha
\int {d^3 x} \left( {\int_{{\bf y}\prime}^{\bf y}\delta ^{\left( 3
\right)} \left( {{\bf x} - {\bf z}} \right)} \right)^2,
\label{NCAE60a}
\end{equation}
\begin{equation}
\left\langle H \right\rangle _\Phi ^{(2)}  = q^2 \beta \int_ {{\bf
y}\prime}^{\bf y} {dz^{\prime i} } \int_ {{\bf y}\prime}^{\bf y}
{dz^i } \frac{{e^{ - m\left| {{\bf z}^\prime - {\bf z}} \right|}
}}{{4\pi \left| {{\bf z}^\prime   - {\bf z}} \right|}},
\label{NCAE60b}
\end{equation}
\begin{equation}
\left\langle H \right\rangle _\Phi ^{(3)}  = q^2 \beta \int_ {{\bf
y}\prime}^{\bf y} {dz^{\prime i} } \int_ {{\bf y}\prime}^{\bf y}
{dz^i } \frac{{e^{ - {\cal M}\left| {{\bf z}^\prime - {\bf z}}
\right|} }}{{4\pi \left| {{\bf z}^\prime   - {\bf z}}
\right|}}.\label{NCAE60c}
\end{equation}
Following our earlier procedure \cite{GaeteGuen,GaeteSpall}, we see
that the potential for two opposite charges located at ${\bf y}$ and
${\bf y^{\prime}}$ takes the form
\begin{equation}
V =  - \frac{{q^2 \left[ {1 + g\left( {{\bf \theta}\cdot {\bf {\cal
B}}} \right)} \right]}}{{4\pi }}\frac{1}{L} + \frac{{q^2 \left( {\bf
{\cal B}} \right)^2 }}{{256\pi M^2 }}\sigma L, \label{NCAE65}
\end{equation}
where
\begin{equation}
\sigma  \equiv g\left( {{\bf \theta}\cdot {\bf {\cal B}}} \right)\ln
\left( {1+\frac{{\Lambda ^2 }}{{m^2 }}} \right) + \left[ {1 +
5g\left( {{\bf \theta}\cdot {\bf {\cal B}}} \right)} \right]\ln
\left( {1 + \frac{{\overline \Lambda  ^2 }}{{{\cal M}^2 }}} \right),
\label{NCAE70}
\end{equation}
while $\Lambda$ and $\overline \Lambda$ are a cuttoff and $|{\bf y}
-{\bf y}^{\prime}|\equiv L$. As already observed, the consistent
limit $\theta\rightarrow0$ has to be taken in Eq. (\ref{NCAE15}) in
order to recover the commutative result.

In summary, we have considered the confinement versus screening
issue for noncommutative axionic electrodynamics, in the case when
there is an external magnetic field. Interestingly, we have obtained
a Cornell-like potential profile. As already expressed, similar
forms of interaction potentials have been reported before in the
context of non-Abelian axionic electrodynamics \cite{GaeteSpall} and
in Yang-Mills theory with spontaneous symmetry breaking of scale
symmetry \cite{GaeteSpall2}. In this way we have provided a new
connection among diverse effective models. The above analysis
reveals the key role played by the noncommutative space in
transforming the Yukawa potential into the Coulombic one.  Also, a
common feature of these models is that the rotational symmetry is
restored in the resulting interaction energy.

Let us also mention here that the static potential profile of these
models is similar to that encountered in a U(1) gauge theory  that
includes the contribution of all topologically nontrivial sectors of
the theory \cite{Kondo}. In this sense these models may be
considered as a physical realization of the topological nontrivial
sectors studied in Ref. \cite{Kondo}. We conclude by noting that
further investigation of the relation between our result and that
encountered in Yang-Mills theory with spontaneous symmetry breaking
of scale symmetry has to be performed, if our analysis represents an
effective approach to non-Abelian gauge theories and especially
quark confinement.

P. G. was partially supported by Fondecyt (Chile) grant 1050546.

\end{document}